\documentstyle[twoside,fleqn,espcrc2,epsf]{article}
\newcommand{\bi}{\begin{itemize}}
\newcommand{\ei}{\end{itemize}}
\newcommand{\beq}{\begin{equation}}
\newcommand{\eeq}{\end{equation}}

\newcommand{\fig}[1]{Fig.~\ref{#1}}

\newcommand{\kk}{$\mathbf k$}
\newcommand{\ee}{$\mathbf E$}
\newcommand{\ez}{$E_z$}
\newcommand{\at}{$A_{\theta}$}
\title{
\hfill\begin{minipage}{0pt}\scriptsize\vspace*{-1.5cm} \begin{tabbing}
\hspace*{\fill} HLRZ1998-53\\ 
\end{tabbing} 
\end{minipage}\\[-8pt]
A Ginzburg-Landau Analysis of the Colour Electric  Flux Tube\thanks{
    Presented by K. Schilling}}
\author{ G.S.~Bali$^{\rm a}$, Ch.~Schlichter$^{\rm c}$, and K.~Schilling$^{\rm b,c}$
 \\ [8pt] 
{\rm $^a$}Institut f\"ur  Physik, Humboldt Universit\"at, Invalidenstr.~110,
D-10115 Berlin, Germany\\
{\rm $^b$}HLRZ, c/o Forschungszentrum J\"ulich, D-52425 J\"ulich,
          and DESY,  Hamburg, Germany \\      
{\rm $^c$}Physics Department, University of Wuppertal,
           D-42097 Wuppertal, Germany\\[8pt]}
\begin{document}
\begin{abstract}
In a simulation of SU(2) gauge theory we investigate, after maximal
 Abelian projection, the dual Maxwell equations for colour field and
 monopole current distributions around a static quark-antiquark pair
 $Q\bar{Q}$ in vacuo. Within the dual superconductor picture
 we carry out a Ginzburg-Landau type analysis of the flux tube
 profile. As a result we can determine the coherence length of the GL
 wave function related to the monopole condensate, $\xi = .25(3) $
 fm, to be compared to the penetration length, $\lambda = .15(2)$ fm (scaled
with the string tension).

\end{abstract}
\maketitle


\section{INTRODUCTION}
 The QCD vacuum can be viewed as a dual superconductor characterized
by a monopole condensate \cite{idea}.  When embedding a static
$Q\bar{Q}$ pair into the vacuum the latter expells by the dual Meissner
effect the colour electric fluxlines, thus giving rise to colour
confinement. The core of the flux tube is just a normalconducting
vortex which is stabilized by solenoidal magnetic supercurrents, \kk,
in the surrounding vacuum.

These features are to hold in some effective Abelian subsector of the
theory.  In a series of previous studies of SU(2) and SU(3) gauge
theories it has indeed been confirmed that the important d.o.f.'s
driving confinement are largely singled out by maximal Abelian gauge
(MAG) projection \cite{mag}, as evidenced e.g. by the string
tension \cite{suzukimag,gauge}.

In this note we will demonstrate that the validity of this dual
superconductivity picture can be extended to colour electric field and
monopole current distributions, \ee\ and \kk, in rather detail.
Refining previous studies along these
lines \cite{haymaker,bari,kanazawa}, we shall perform an unbiassed
analysis of the Ginzburg-Landau (GL) wave function \cite{GL}, $f$, which can
be extracted in a sequence of steps from the profile of the flux tube
in the transverse center plane (CP) to $Q\bar{Q}$, as seen on the
lattice. $f$ carries a second length scale (in addition to the dual
photon mass, $\lambda$), in form of the coherence length, $\xi$, of
the condensate; it is the counterpart to the Abelian Higgs mass in the
effective GL approach.

\section{EVIDENCE FOR  DUAL THEORY}
\begin{figure}[htb]
\vspace{-1.6cm}
\epsfxsize=8.0cm\epsfbox{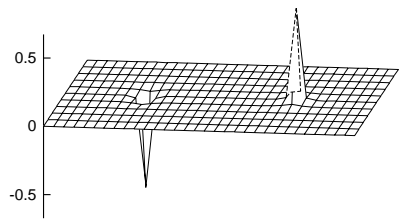}
\vskip -2cm
\hskip -6mm
\caption{div $\mathbf{E}$ is strictly localized to the position of $Q$
  and $\bar{Q}$, at separation $r\approx 1.2$ fm.}.
\label{divE}
\vspace{-0.9cm}
\end{figure}

\begin{figure}[htb]
\vspace{-0.6cm}
\epsfxsize=8.0cm\epsfbox{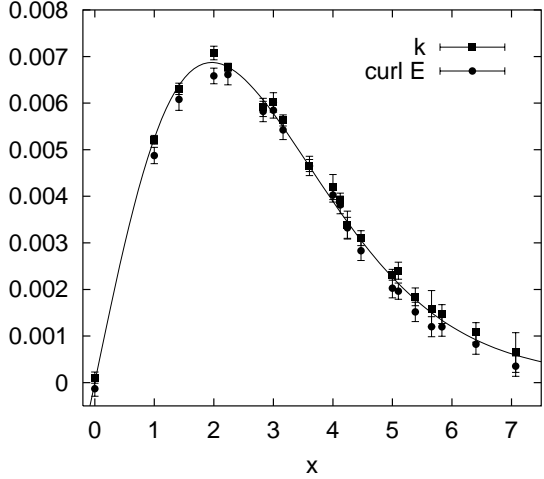}
\vskip -1cm
\hskip -6mm
\caption{Test of the dual Amp\`ere Law, $\mathbf{k} = \mbox{curl}
  {\mathbf E}$.}
\label{ampere}
\vspace{-0.9cm}
\end{figure}

It is crucial to start out by establishing the validity of the dual
Maxwell theory. We create SU(2) field configurations in the usual way,
go through MAG projection into the Abelian world \cite{gauge} and
determine correlators between large Wilson loops and plaquettes
therein \cite{flux}, while \kk\ is defined  according to the
prescription of our local host \cite{degrand}.  In \fig{divE} we verify
that source and sink of \ee\ are strictly localized to the positions
of $Q$ and $\bar{Q}$!
 \fig{ampere} illustrates that the data on CP transverse
distributions (plotted versus radial distance in lattice units, $x$, out of core) provides
impressive support of  the dual Amp\`ere Law. 

\section{GINZBURG-LANDAU ANALYSIS}
Being nonlinear, the GL equations \cite{GL} are not at all trivial to solve
on a grid:
\begin{eqnarray}
\label{GL_f}
f(x) &=& f(x)^3 + \xi^2
\left[F(x)^2
-\frac{1}{x}\frac{d}{dx}\left(x\frac{d}{dx}\right)\right]f(x), \nonumber\\
\label{GL_k} 
k_\theta(x) &=&
\frac{f(x)^2}{\lambda^2}\frac{\Phi}{2\pi x}F(x),\\
\mbox{with} \nonumber\\
F(x) &=& \left(\frac{1}{x}-\frac{2\pi
A_\theta(x)}{\Phi}\right) \quad .
\end{eqnarray}
Our approach is to determine $f$ directly from the electric vector
potential \at (through integration of ${\mathbf E}=\mbox{curl}{\mathbf A}$)
and $k_{\theta}$ via the second GL equation (GLII). For this purpose we employ a parametrization for $E_z$ that respects the boundary conditions
of $f$ (BCf)
which are  rather involved, as they imply the value 
\begin{figure}[htb]
\vspace{-0.6cm}
\epsfxsize=8.0cm\epsfbox{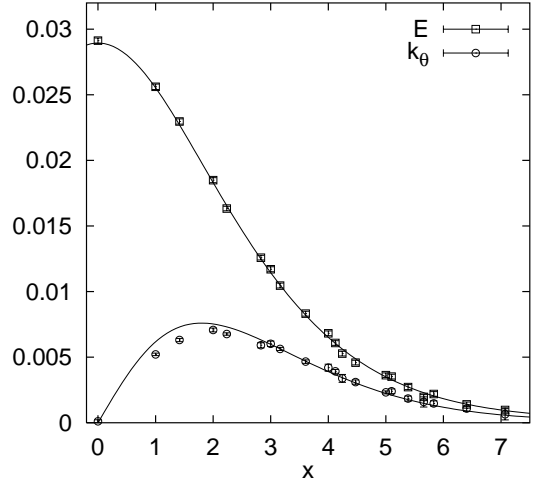}
\vskip -1cm
\hskip -6mm
\caption{Fitting the azimuthal magnetic supercurrent and longitudinal colour electric field.}.
\label{fits}
\vspace{-0.9cm}
\end{figure}
of $E_z(0)$ \cite{kyoto}.

Given the functional form of \ez\ the dual Amp\`ere Law induces a
parametrization for $k_{\theta}$ as well, which at the end leads us to
a 4 parameter ansatz to fit both distributions simultaneously.  For a
discussion of the details we refer to ref. \cite{kyoto}. The two
interesting physics parameters resulting from the joint fit to \ez\
and $k_{\theta}$, namely the total colour electric flux $\Phi$
(expected to have the value {\it one}) and the penetration length
$\lambda$ come out to be $ \Phi=1.08(2),\quad \lambda=1.84(8)$, in
consistency with the value from a deep London limit analysis, $\lambda
= 1.82(7)$ \cite{kyoto}. The simulations are performed on a $32^4$
lattice at $\beta = 2.5115$, the lattice spacing being
$a\approx 0.081$ fm~\cite{flux}.
 
\begin{figure}[htb]
\vspace{-0.6cm}
\epsfxsize=8.0cm\epsfbox{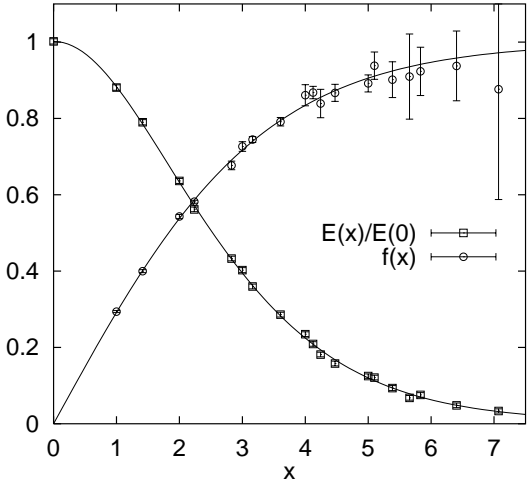}
\vskip -1cm
\hskip -6mm
\caption{GL wavefunction and colour electric field in transverse profile.}.
\label{boundary}
\vspace{-0.9cm}
\end{figure}

 The quality of the fit is illustrated in \fig{fits}, where we see
nice agreement apart from some lattice artifacts in $k$ in the small
$x$ regime, $x < 2.2$\footnote{We meet non-integer lattice spacings as
we use off-axis separations!}. As a result we would expect to be
able to determine $f$ from solving  GLII
outside this region. The result is shown in \fig{boundary} where we
plotted the CP transverse distribution of the wave function, together
with the profile of the longitudinal field $E_z(x)$. Notice that we
attain reasonable statistical accuracy on $f$, up to about 4.2 lattice
spacings out of the vortex core. Beyond this point, we are loosing
overly in sensitivity for $f$. Nevertheless one can observe $f$ to
approach asymptotia, $f(\infty) = 1$, fairly rapidly.

\begin{figure}[htb]
\vspace{-0.6cm}
\epsfxsize=8.0cm\epsfbox{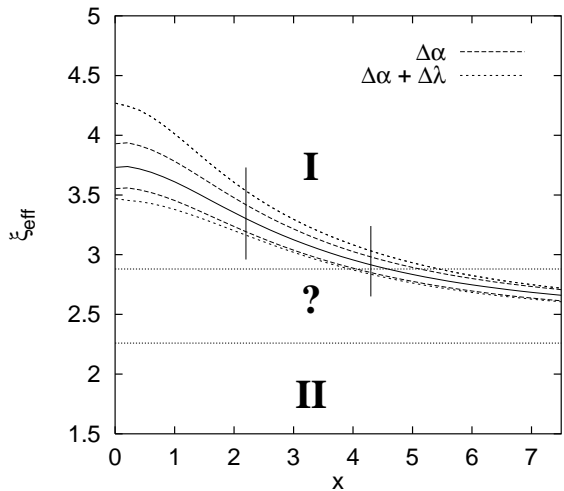}
\vskip -1cm
\hskip -6mm
\caption{Effective $\xi$ with error bands related to uncertainties
  in
$\alpha$ and $\lambda$.}.
\label{effective}
\vspace{-0.9cm}
\end{figure}

In order to finally extract the quantity of interest, $\xi$, it is
useful to parametrize $f$ with a functional form that again obeys
BCf. To this end  we try the usual
ansatz $  f(x) = \tanh(x/\alpha) $,
which is very successful, as evidenced in \fig{boundary}, with
$\alpha = 3.33(5)$, but $\lambda = 1.62(2)$\footnote{Note
that $\alpha$ is not identical to the coherence length as sometimes
assumed.}.  Now that we have a good continuum form of $f(x)$ from GLII
at our disposal, we can plug it into  GLI  and solve
the latter in terms of an effective function $\xi (x)$. To the extent
that this function proves to be constant we can claim to have arrived
at a consistent determination of the coherence length, $\xi$. In
\fig{effective} we exhibit the outcome: indeed $\xi$ shows only a
rather mild ($\pm 10 \%$) $x$ dependency within our window of
observation $2.2 < x < 4.2$. Finally, translating the remaining
variation into the systematic error, we can quote the value 
$\xi = 3.10^{+43}_{-35}$.
\section{DISCUSSION}
If one includes the uncertainties in the determination of $\alpha$ and
$\lambda$, one obtains the error bands for $\xi$ as indicated in
\fig{effective}. A conservative estimate 
of the uncertainty in the determination of the penetration length, 
$\lambda = 1.84^{+20}_{-24}$,  
 blurs the estimated
transition point between type I and type II superconductivity, $\kappa
= \lambda/\xi = 1/\sqrt{2}$ into an error interval, as indicated in
\fig{effective} by the symbol ``?''. We conclude that there is {\it
weak evidence} for type I superconductivity.

It would be worthwhile to confirm this tentative conclusion by a
scaling analysis and to study interaction among vortices by putting
more than one of them into the vacuum.

{\bf Acknowledgements. }
G.S.~B. thanks DFG for support under  grant Ba 1564/3.


\end{document}